# Reactive Fluid Ferroelectrics: A Gateway to the Next Generation of Ferroelectric Liquid Crystalline Polymer Networks

Stuart R. Berrow [1], Jordan Hobbs [1], and Calum J. Gibb [2*]

[1]School of Physics and Astronomy, University of Leeds, Leeds, UK, LS2 9JT

[2]School of Chemistry, University of Leeds, Leeds, UK, LS2 9JT

*Author for correspondence: c.j.gibb@leeds.ac.uk

**Abstract**

Herein we report the first examples of reactive mesogenic materials (RMs) which exhibit fluid ferroelectric order based on the recently discovered ferroelectric nematic ($N_F$) phase. We term these materials $N_F$ RMs and they provide the first steps towards the next generation of ferroelectric liquid crystalline polymer networks. We report the chemical synthesis and characterisation of the liquid crystalline properties of these materials, demonstrating that they have the lowest longitudinal molecular dipole moments ($\mu$) of any reported $N_F$ material of 7.39 D. We go on to demonstrate a potential use case of this new class of reactive material through the polymer stabilisation of a matrix which exhibits the $N_F$ phase, increasing the phase range of the ferroelectric phase from 75 °C to 120 °C. The $N_F$ RMs reported herein are an exciting step forward in ferroelectric liquid crystal research, demonstrating that reactive $N_F$ materials are achievable, allowing for the future development of liquid crystalline ferroelectric networks, elastomers and polymers.

Reactive mesogens (RMs) are functional materials displaying liquid crystalline (LC) mesophases that possess reactive groups, allowing for the formation of polymers, elastomers and network structures with liquid crystalline order. The first RM was synthesised in 1989 [1] as a means of protecting optical fibres, but it was not long until applications in the fabrication of liquid crystal displays (LCDs) emerged, leading to far reaching implications in the multibillion-dollar LCD industry [2–5]. Over the past 35-years, the application of reactive mesogenic materials has continued to grow giving rise to new fields of study, such as: liquid crystal elastomers (LCEs) [6,7], liquid crystal templating [8], polymer dispersed LC networks [9], and the stabilisation of LCs for device applications – commonly referred to as polymer stabilised liquid crystals (PSLCs) [10].

The recent discovery of the ferroelectric nematic ($N_F$) phase, which combines nematic orientational order and fluidity with almost perfect parallel polar order of its constituent molecules [11,12], has garnered significant scientific interest due to its potential applications such as fluid piezoelectrics [13], tuneable lasers [14] and reflectors [15], non-linear electrooptical devices [16,17], and the generation of entangled photon pairs [18]. The potential scope of the $N_F$ phase would be further increased through its incorporation into network structures. Whilst some work has already been done on macromolecular $N_F$ materials [19–21], a polymer network with inherent $N_F$ order has yet to be reported but would be desirable [22]. An obvious direction would be polymer stabilisation of the $N_F$ phase. So far, it has been reported that ferroelectric nematogens have poor miscibility with the non-mesogenic network structures typically applied, resulting in phase separation [23]. Some successes have been achieved using blue-phase networks [23], and optically isotropic polymer stabilised materials [24,25]. An $N_F$ exhibiting RM would be desirable as it would not only improve miscibility with $N_F$ phases for polymer stabilisation but would also give a pathway to new polymer networks that are polar in their own right. To date, there is no such RM reported within the literature and so we elected to design and synthesise a ferroelectric nematogen with reactive functionality.

As research into the $N_F$ phase is still largely in its infancy, there are only a few structure spaces currently known to promote the formation of the phase [26–28]. 4-(difluoro(3,4,5-trifluorophenoxy)methyl)-1,1'-biphenyl) is a common molecular feature present in several materials exhibiting the $N_F$ [29,30] (and other polar mesophase [31,32]) and so we chose this as the basic unit for our RMs. Coupling this to a basic acrylate unit afforded the four reactive materials **1-4 (Table 1)**. Of the four RMs prepared, two of the structures, **2** and **3**, exhibit the $N_F$ phase. For the sake of brevity, we will focus our discussion on **2** while data for the remaining materials are provided in the ESI to this article.

**Table 1.** Transition temperatures and their associated enthalpy changes [in brackets] for **1-4**. Their longitudinal molecular dipole moments (μ / D) (at the DFT:B3LYP-GD3BJ/cc-pVTZ level [35–38]) are also given in the left-hand column. Crystallisation temperatures (Crystal) recorded on cooling.

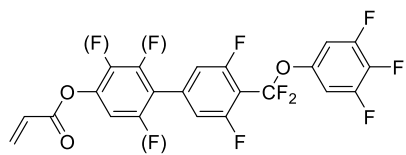

| Compound | Melt T / °C [ΔH / KJ mol⁻¹] | Crystal T / °C [ΔH / KJ mol⁻¹] | SmA_F - N_F T / °C [ΔH / KJ mol⁻¹] | N_F-I T / °C [ΔH / KJ mol⁻¹] | N-I T / °C [ΔH / KJ mol⁻¹] |
|---|---|---|---|---|---|
| **1** μ = 6.87 D | 73.9 [25.8] | 53.3 [21.7] | - | - | 77.1 [0.5] |
| **2** μ = 7.39 D | 81.6 [15.2] | 23.4 [14.8] | 27.7 [0.5] | 84.3 [2.9] | - |
| **3** μ = 7.98 D | 112.4 [21.6] | 99.7 [25.4] | - | 116.2 [5.7] | - |
| **4** μ = 7.13 D | 103.0 [32.4] | 90.6 [32.2] | - | - | - |

Briefly, before beginning the discussion of **2**, it is of note that all four homologues have relatively low longitudinal molecular dipole moments (μ) (**Table 1**) compared to the vast majority of reported ferroelectric nematogens [26,27]. Although the importance of large values of μ with regards to the formation of the polar phases is currently debated[39–43] all existing reported N_F materials possess dipole moments greater than 8.5 D with the exceptions generally belonging to polar smectic materials[41–43]. For example, the archetypal materials RM734 and DIO have dipole moments of ~11 D and ~9 D, respectively [26]. Here **2** and **3** possess dipole moments of 7.39 D and 7.98 D, respectively, making them the lowest of any reported material exhibiting the N_F phase to date. Empirically, this reinforces the emerging trend that the magnitude of μ may not be a limiting factor for thermal stability of the N_F phase for all materials [41].

The phase transition properties of the **2** were determined from polarized optical microscopy (POM), differential scanning calorimetry (DSC) and current reversal measurements (experimental details included in the ESI). When viewed between untreated glass, cooling **2** from the isotropic (I) phase sees the N_F phase form from small droplets (**Fig. 1a [left]**), characteristic of the N_F phase [42]. Further cooling sees the coalescence of the droplets into a banded texture, synonymous with the assignment of the N_F phase (**Fig. 1a [centre]**). Rapid cooling to 25 °C showed a further transition to a phase with a blocky mosaic texture associated with the SmA_F phase (**Fig. 1a [right]**) [43,44]. Attempts to confirm this assignment via X-ray scattering were unsuccessful due to crystallisation of the sample during measurements therefore the assignment of the SmA_F phase here is tentative. Transition temperatures for these phases were obtained by DSC, with transitions appearing as first-order peaks in the relevant DSC thermograms (**Table 1, Fig. 1b, and Fig. S5**). Ferroelectric order was confirmed via current response measurements (**Fig. 1c and Fig. S7**), where a single polarisation reversal peak was observed.

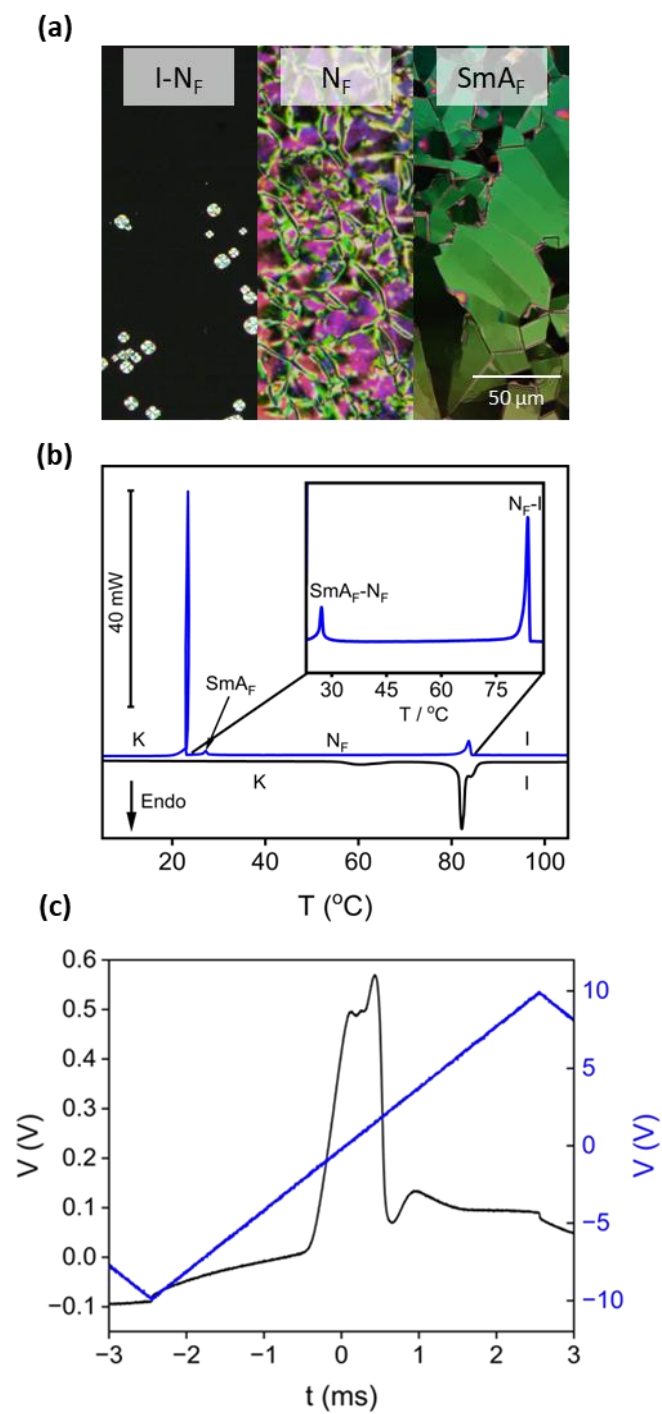

**Figure 1.** (a) POM micrographs taken for **2** depicting **[left]** the I-$N_F$ phase transition where the $N_F$ phase forms from droplets, **[centre]** the banded texture of the $N_F$ phase, and **[left]** the blocky mosaic structures associated with the $SmA_F$ phase. Images were taken within thin cells with no alignment layer at 84 °C, 50 °C, 25 °C, respectively; (b) DSC thermogram of **2** showing heating and cooling traces; and (c) current response trace measured for **2** measured at 100 Hz in the $N_F$ phase at 80 °C.

Presently, polymer stabilisation of the $N_F$ phase has not yet been reported. We elected to demonstrate the potential viability of our $N_F$ RMs by polymer stabilising a known, unreactive $N_F$ material. We envisaged that if the network contains repeat units that exhibit the $N_F$ phase, there will be two advantages. Firstly, the issue of poor affinity will be mitigated by the ideal mixing behaviour that occurs between $N_F$ molecules in mixtures [39,45]. Secondly, it is likely that a key component for the formation of a polar network is to polymerise in a polar phase. The addition of a non-$N_F$ RM would cause a significant reduction in the $N_F$ transition temperature preventing polymerisation in the $N_F$ phase. By using an $N_F$ RM, this is avoided completely.

For our test, we elected to stabilise the phase behaviour of the simple mixture **F7 (Fig. 2a)** [31], a simple binary mixture of two polar materials which exhibits the $N_F$ phase from room temperature up to 95 °C before transitioning to the splay nematic ($N_S$) phase (an antiferroelectric nematic phase sometimes referred to as $N_X$, $N_{AF}$ or $SmZ_A$ [46,47]). A series of mixtures containing varying quantities of **2** (**Fig. 2b**), the photoinitiator **MBF** (**Fig. 2c**), and the commercially available crosslinker **RM82** (**Fig. 2d**) in **F7** were produced, ensuring the concentrations of **F7** and **MBF** were kept constant (**Table 2**). Prior to polymerisation, each mixture was examined using the same physical characterisation techniques as materials **1-4**, confirming that the $N_F$ phase was exhibited by each mixture (**Fig. 3 a-c, Fig. S8** and **9**). For **PS3** and **PS4** the transition temperatures associated with the $N_F$ phase fall dramatically compared to **F7** due to the increased quantity of apolar **RM82** destabilising polar order.

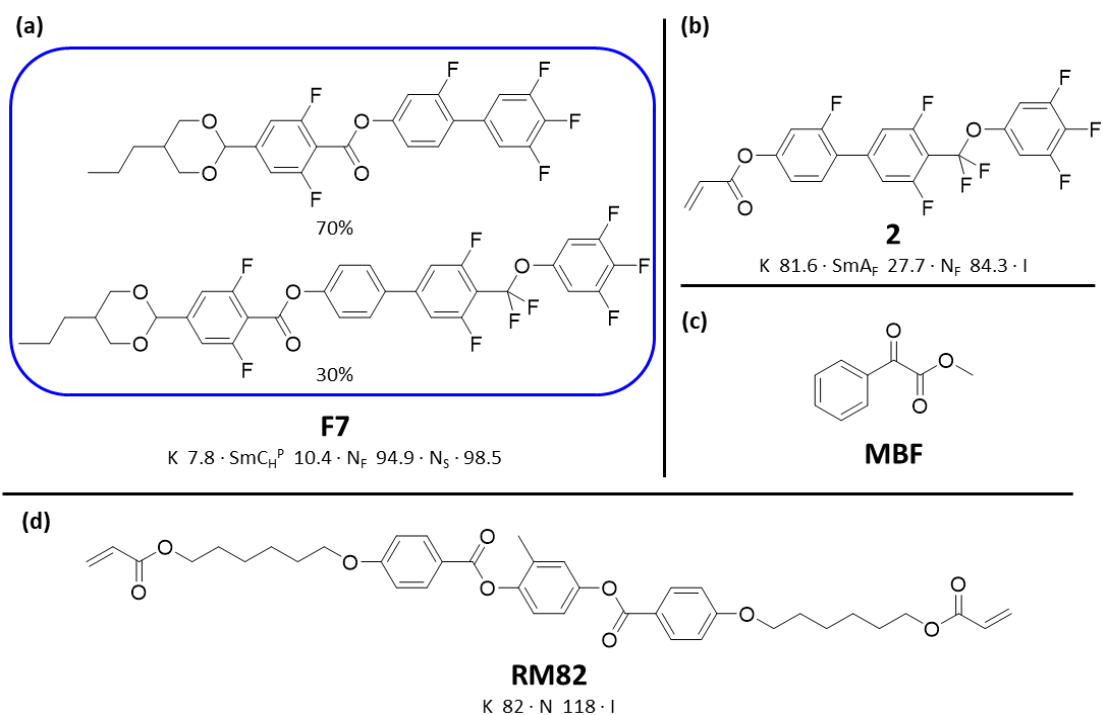

**Figure 2**. The chemical structures and their transition temperatures, where appropriate, used in the formulation of mixtures **PS1-4**.

**Table 2.** The compositions of the polymer stabilised mixtures **PS1-4**, and the phase transition temperatures for the K-$N_F$ and $N_F$-$N_S$ phase transitions pre- and post-polymerisation.

| Mixture Name | 2 (wt %) | RM82 (wt %) | MBF (wt %) | F7 (wt%) | Pre-polymerisation $T_{K-NF}$ (°C) | Pre-polymerisation $T_{NF}$ (°C) | Post-Polymerisation $T_{K-NF}$ (°C) | Post-Polymerisation $T_{NF}$ (°C) |
|---|---|---|---|---|---|---|---|---|
| PS1 | 8.65 | 1.25 | 0.1 | 90 | - | 89.9 | - | 89.8 |
| PS2 | 7.4 | 2.5 | 0.1 | 90 | - | 86.6 | - | 90.6 |
| PS3 | 4.9 | 5.0 | 0.1 | 90 | 24.1 | 78.5 | - | 86.3 |
| PS4 | 0 | 9.9 | 0.1 | 90 | 22.5 | 54.5 | 20.7 | 86.4 |

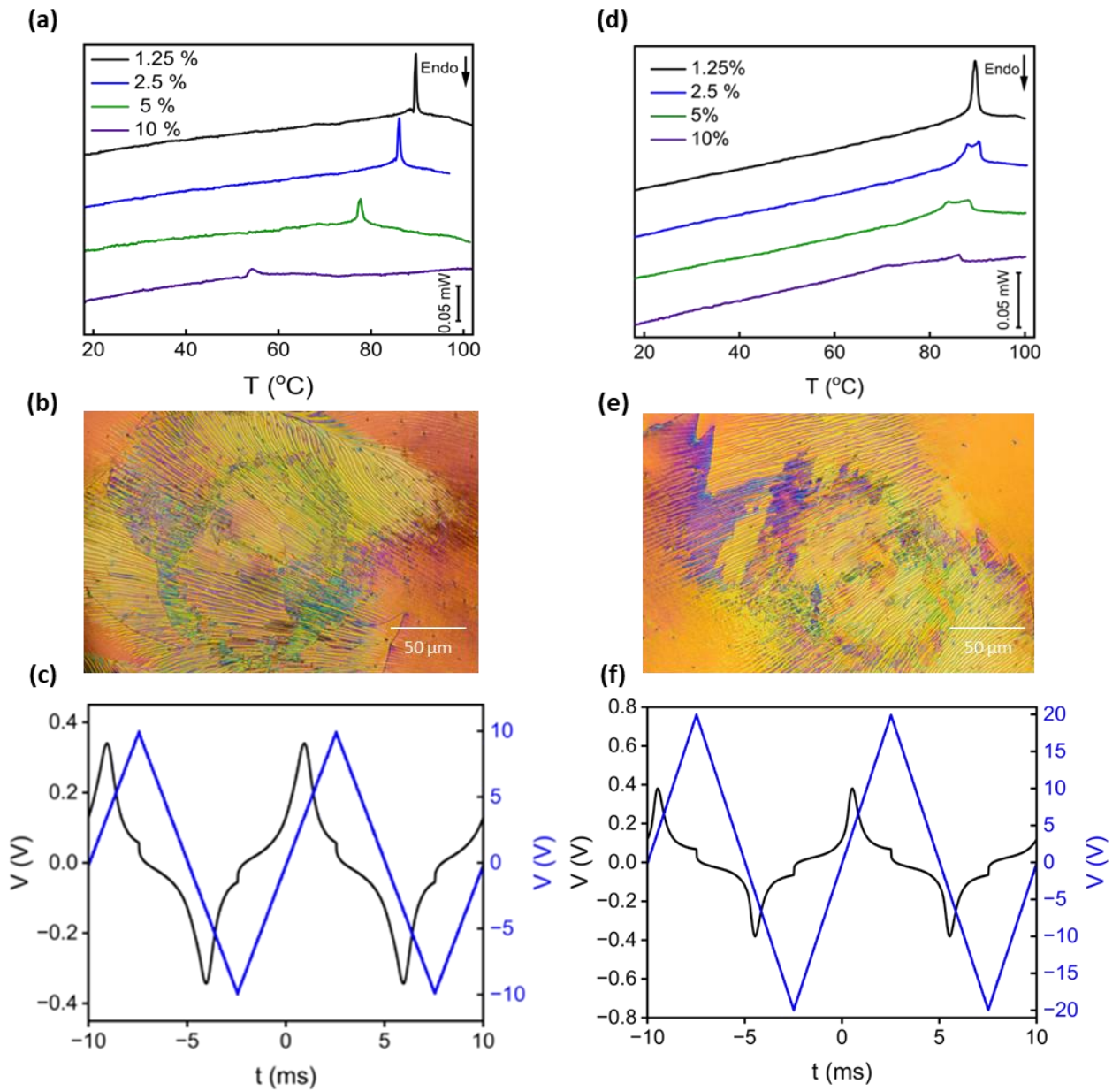

**Figure 3.** (a) DSC thermograms of **PS1-4**, (b) POM micrograph of **PS2** at 40 °C, and (c) current response trace for **PS2** measured at 100 Hz in the $N_F$ phase at 30 °C pre-polymerisation. (d) DSC thermograms of **PS1-4**, (e) POM micrograph of **PS3**, and (f) current response trace for **PS3** measured at 100 Hz in the $N_F$ phase at 30 °C post-polymerisation.

Post-polymerisation using UV light (365 nm (2.5 W cm$^{-2}$, 20 min), the N$_F$ phase is retained for all four mixtures and an increase in the temperature stability of the N$_F$ phase is observed for **PS2-4** with the transition temperature of the N$_F$ phase in **PS1** remaining comparable to the uncured sample (**Table 2, Fig. 3d**). This is due to the low concentration of **RM82** present in this mixture indicating a minimum degree of crosslinking is required for stabilisation. The characteristic banded texture of the N$_F$ is retained post-polymerisation by all mixtures (**Fig. 3e**) indicating that this polymer stabilisation mixture does not prevent the separation of the N$_F$ texture into domains of alternative polarisation. A stronger electric field is required to facilitate the switching of the polymerized materials (**Fig. 3f**) likely due to an increased elastic deformation cost which could be due to the creation of stronger polar anchoring [48] by the now polymerised network or some other restriction imposed by the resulting network structures. There is also likely to be an element of increased rotational viscosity.

The largest degree of stabilisation in the N$_F$ phase is seen for **PS4**, a mixture containing no **2** and, to a first approximation, it appears that an N$_F$ RM is not required to successfully stabilise the N$_F$. However, both a significant broadening of the first order peak associated with the N$_F$ transition (**Fig. 3**) and the cold crystallisation of the sample on re-heating (**Fig. S8**) are evidence of significant phase separation [51]. Although the degree of stabilisation in **PS2** and **PS3** is lower than that observed in **PS4**, there is clearly significantly less de-mixing (**Fig. S8d**). Whilst incorporating a N$_F$ RM into the mixture reduces the effect of the stabilisation, its presence is clearly beneficial, enhancing the miscibility between the network and the matrix.

To summarise, we have reported the synthesis of the first reactive mesogen materials to natively display the N$_F$ phase which we term N$_F$ RMs. We then show a potential use case for this new class of functional materials through the polymer stabilisation of the matrix **F7** [31]. When mixed in conjunction with a small amount of crosslinker (**RM82**) and photoinitiator (**MBF**), we demonstrate the successful stabilisation of the N$_F$ matrix increasing the effective N$_F$ phase range from 75 °C to 120 °C. By using inherently ferroelectric RMs we act to mitigate the phase separation issues that have been problematic in the previously attempted network structures [23]. We expect that N$_F$ RMs may be pivotal in the production of new ferroelectric liquid crystalline networks, elastomers and polymers, thereby increasing the possible use cases of the N$_F$ phase.

## Data availability

The data associated with this paper are openly available from the University of Leeds Data Repository at https://doi.org/10.5518/1635.

## Author Contribution Statement

All authors discussed all results, and the manuscript was subsequently written, reviewed, and edited with contributions from all authors.

## Competing interests

Authors declare that they have no competing interests.

## Acknowledgements

The authors would like to thank Prof. Helen F. Gleeson and Dr. Richard J. Mandle for useful discussions, support and guidance. The authors would like to acknowledge funding from the Engineering and Physical Sciences Research Council, Grant Number EP/V054724/1 and, UKRI for funding via a Future

Leaders Fellowship, grant number MR/W006391/1. Computational work was performed on ARC4, part of the high-performance computing facilities at the University of Leeds.

# Reactive Fluid Ferroelectrics: A Gateway to the Next Generation of Ferroelectric Liquid Crystalline Networks

S. R. Berrow[1], J. Hobbs[1], and C. J. Gibb[2*]

[1]School of Physics and Astronomy, University of Leeds, LS2 9JT

[2]School of Chemistry, University of Leeds, LS2 9JT

Correspondence: C.J.Gibb@leeds.ac.uk

## Electronic Supplementary Information

**Supporting Information available**: Experimental information, synthetic procedures for the synthesis of all monomers and intermediates, NMR spectra, Characterisation of monomers (DSC, POM and current reversal) and polymer stabilisation studies (procedure, DSC and current reversal).

## Table Of Contents

# Experimental Information

## Materials

Chemicals were purchased from commercial suppliers (Fluorochem, Merck, ChemScene, Ambeed) and used as received. Solvents were purchased from Merck and used without further purification. Reactions were performed in standard laboratory glassware at ambient temperature and atmosphere (unless otherwise stated) and were monitored by TLC with an appropriate eluent and visualised with 254 nm light.

## Flash Chromatography

Flash chromatography was performed on a Combiflash NextGen 300+ system (Teledyne Isco) using silica as a stationary phase, an appropriate mobile phase as specified in the experimental procedure, and detection in the 200-800 nm wavelength range.

## Structural Analysis

Nuclear magnetic resonance (NMR) spectra were recorded using either a Bruker AVANCE III (400 MHz) NMR spectrometer (Bruker UK Ltd., Coventry, UK) or a Bruker AV4 NEO 11.75T (500 MHz) spectrometer, with spectra collected at 298 K and referenced to TMS. NMR spectra were viewed and analysed using MNova NMR software.

## Thermal Analysis

Differential scanning calorimetry (DSC) measurements were performed using a TA Instruments Q2000 DSC instrument (TA Instruments, Wilmslow UK), equipped with a RCS90 Refrigerated cooling system (TA Instruments, Wilmslow UK). The instrument was calibrated against an Indium standard, and data were processed using TA Instruments Universal Analysis Software. Samples were analysed under a nitrogen atmosphere, in hermetically sealed aluminium TZero crucibles (TA Instruments, Wilmslow, UK) and subjected to three analysis cycles. In all cases, samples were subject to heating and cooling at a rate of 10 K min$^{-1}$. Phase transition temperatures were measured as onset values on cooling cycles for consistency between monotropic and enantiotropic phase transitions, while crystal melts were obtained as onset values on heating.

## Optical Microscopy

Polarised light optical microscopy (POM) was performed using a Leica DM2700P polarised light microscope (Leica Microsystems (UK) Ltd., Milton Keynes, UK), equipped with 10x and 50x magnification, and a rotatable stage. A Linkam TMS 92 heating stage (Linkam Scientific Instruments Ltd., Redhill, UK) was used for temperature control, and samples were studied sandwiched between two untreated glass coverslips. Images were recorded using a Nikon D3500 Digital Camera (Nikon UK Ltd., Surbiton, UK), using DigiCamControl software.

## Electronic Structure Calculations

Electronic structure calculations were performed using Gaussian G16 revision C.02 [1] and with a B3LYP-GD3BJ/cc-pVTZ [2–5] basis set. Obtained structures were verified as a minimum from frequency calculations. Following geometry optimisation, a frequency calculation at the same level was used to confirm the geometry to be at a minimum.

## Measurement of Spontaneous Polarization (P$_S$)

Spontaneous polarisation measurements are undertaken using the current reversal technique [6,7]. Triangular waveform AC voltages are applied to the sample cells with an Agilent 33220A signal generator (Keysight Technologies), and the resulting current outflow is passed through a current-to-voltage amplifier and recorded on a RIGOL DHO4204 high-resolution oscilloscope (Telonic Instruments Ltd, UK). Heating and cooling of the samples during these measurements is achieved with an Instec HCS402 hot stage controlled to 10 mK stability by an Instec mK1000 temperature controller. The LC samples are held in 4µm thick cells with no alignment layer, supplied by Instec. The measurements consist of cooling the sample at a rate of 1 Kmin$^{-1}$ and applying a set voltage at a frequency of 10 Hz. The voltage was set such that it would saturate the measured P$_S$ and was determined before final data collection.

There are three contributions to the measured current trace: accumulation of charge in the cell (I$_c$), ion flow (I$_i$), and the current flow due to polarisation reversal (I$_p$). To obtain a P$_S$ value, we extract the latter, which manifests as one or multiple peaks in the current flow, and integrate as:

$$P_S = \int \frac{I_p}{2A} dt \quad \textbf{(1)}$$

where A is the active electrode area of the sample cell. For the N, N$_X$ and, to a lesser extent, the N$_F$ phase, significant amounts of ion flow is present. For materials that showed a paraelectric N phase followed by the anti-ferroelectric N$_X$ phases, the N phase always showed some pre-transitional polarisation as well as the significant ion flow mentioned previously. The P$_S$ of the N$_X$ phases was obtained by integrating the peak least affected by ion flow and then doubled to get the total area under both peaks [8].

## Synthetic Procedures

The synthetic route to the four monomers is displayed in **Scheme S1**. The synthesis of intermediate **I1** [9] and the four phenolic reagents (**I2-5**) [9,10] are reported elsewhere.

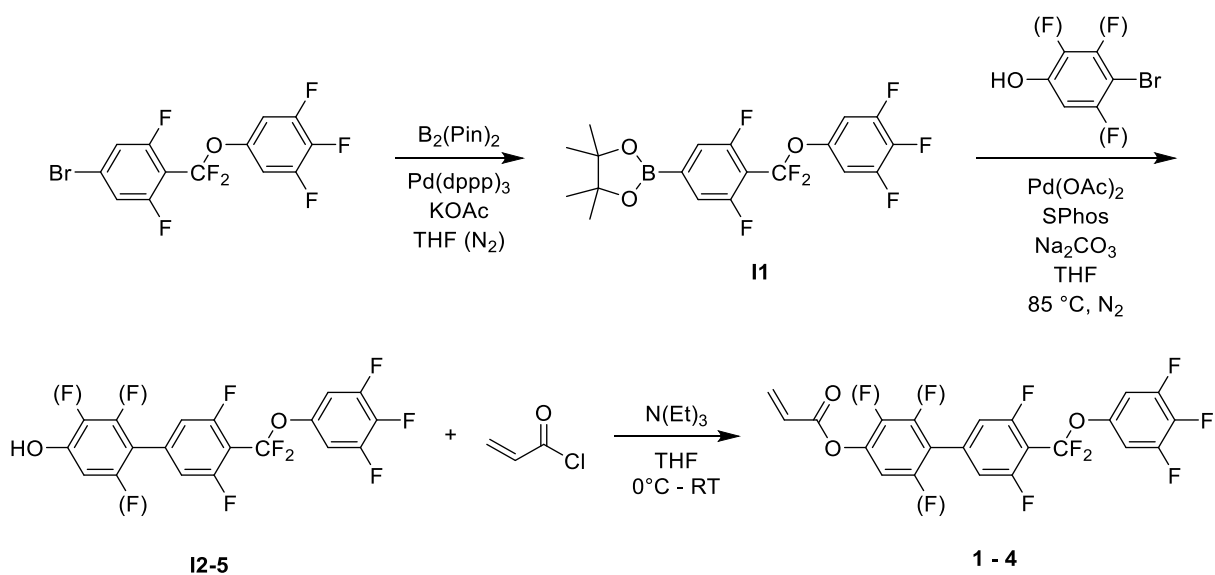

**Scheme S1** – *The synthetic pathway employed in the synthesis of monomers **1-4**.*

## General Procedure for Synthesis of Monomers 1-4

A round bottomed flask was charged with the appropriate phenol (1 mmol, 1.0 eq), triethylamine (1.5 mmol, 0.21 mL, 1.5 eq) and THF (conc. ~ 0.2 M) and the flask cooled over an ice-water bath. To the reaction mixture was added acryloyl chloride (0.12 mL, 1.5 mmol) dropwise of a period of 15 mins, before the reaction mixture was allowed to warm to room temperature and stirred overnight. The solvent was removed under reduced pressure, and the resulting material dissolved in chloroform (100 mL). The organic layer was washed with aqueous NaHCO$_3$ solution (1M) (2x100 mL) and brine (saturated) (100 mL), and the organic layer dried over magnesium sulphate. The solvent was removed under reduced pressure, and the crude solid recrystallized from isopropanol, to yield the products as colourless crystals.

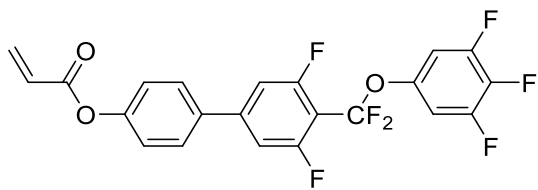

*4-acryloxy-(3',5'-difluoro-4'-(difluoro(3,4,5-trifluorophenoxy)methyl)-1,1'-biphenyl)* **(1)**

Yield: (Colourless crystals) 180 mg, 40%

$R_F$ (50:50 Hexane:DCM): 0.47

$^1$H NMR (501 MHz, CDCl$_3$) (δ): 7.51 (ddd, J = 8.7, 2.0, 2.0 Hz, 2H, Ar-**H**), 7.19 (ddd, J = 8.7, 2.0, 2.0 Hz, 2H, Ar-**H**), 7.12 (d$_{apparent}$, J = 10.2 Hz, 2H, Ar-**H**), 6.91 (dd, J = 7.9, 5.8 Hz, 2H, Ar-**H**), 6.57 (dd, J = 17.3, 1.2 Hz, 1H, **H**C=CH$_{trans}$H$_{cis}$), 6.27 (dd, J = 17.3, 10.5 Hz, 1H, HC=C**H**$_{trans}$H$_{cis}$), 5.99 (dd, J = 10.4, 1.2 Hz, 1H, HC=CH$_{trans}$**H**$_{cis}$).

$^{13}$C{$^1$H} NMR (126 MHz, CDCl$_3$) (δ): 164.46, 160.41 (dd, J = 257.6, 6.1 Hz), 152.20, 151.20 (ddd, J = 250.7, 10.7, 5.0 Hz), 146.20 (t, J = 10.6 Hz), 144.79 (td, J = 10.6, 4.0 Hz), 138.59 (dt, J = 250.4, 15.2 Hz), 135.27, 133.31, 128.28, 127.76, 119.32 (d, J = 266.1 Hz), 118.26, 116.24, 115.43, 111.13 (dd, J = 23.4, 3.5 Hz), 108.95 – 108.04 (m$_{apparent}$), 107.59 (dd, J = 18.3, 5.6 Hz).

$^{19}$F NMR (376 MHz, CDCl$_3$) (δ): -61.66 (t, J = 26.2 Hz, 2F, O-C**F$_2$**-Ar), -110.11 (td, J = 26.2, 11.1 Hz, 2F, Ar-**F**), -132.45 (dd, J = 20.6, 8.6 Hz, 2F, Ar-**F**), -163.14 (tt, J = 20.7, 5.9 Hz, 1F, Ar-**F**).

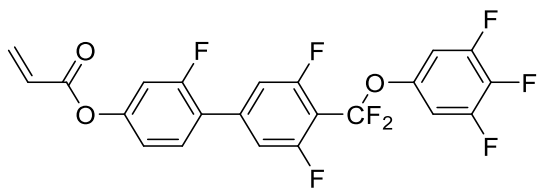

*4-acryloxy-2-fluoro-(3',5'-difluoro-4'-(difluoro(3,4,5-trifluorophenoxy)methyl)-1,1'-biphenyl)* **(2)**

Yield: (Colourless crystals) 169 mg, 36%

$R_F$ (50:50 Hexane:DCM): 0.51

$^1$H NMR (501 MHz, CDCl$_3$) (δ): 7.46 (t, J = 8.6 Hz, 1H, Ar-**H**), 7.20 (d$_{apparent}$, J = 10.6 Hz, 2H, Ar-**H**), 7.12 – 7.06 (m, 2H, Ar-**H**)*, 6.99 (dd, J = 7.9, 5.8 Hz, 2H, Ar-**H**), 6.66 (dd, J = 17.3, 1.1 Hz, 1H, HC=C-**H**$_{trans}$H$_{Cis}$), 6.34 (dd, J = 17.3, 10.5 Hz, 1H, **H**C=CH$_{trans}$H$_{cis}$), 6.09 (dd, J = 10.4, 1.1 Hz, 1H HC=C-H$_{trans}$**H**$_{Cis}$).
*overlapping peaks

$^{13}$C{$^1$H} NMR (126 MHz, CDCl$_3$) (δ): 164.01, 161.10 (dd, J = 257.7, 6.0 Hz), 159.35 (d, J = 250.9 Hz), 152.15 (dd, J = 11.0, 4.0 Hz), 150.16 (dd, J = 10.9, 5.2 Hz), 144.74 (td, J = 12.3, 3.9 Hz), 140.89 (t, J = 11.1 Hz), 138.12 (dt, J = 251.1, 15.2 Hz), 133.82, 130.65 (d, J = 3.8 Hz), 127.43, 123.33 (d, J = 12.7 Hz), 122.39, 120.27, 118.41 (d, J = 3.6 Hz), 118.15, 113.22 (dt, J = 23.8, 3.6 Hz), 110.87 (d, J = 25.7 Hz), 109.52 – 108.60 (m$_{apparent}$), 107.62 (dd, J = 19.0, 6.0 Hz).

$^{13}$C{$^1$H}{$^{19}$F} NMR (126 MHz, CDCl$_3$) (δ): 163.87, 160.55, 159.94, 159.51, 158.47, 152.03, 144.60, 140.76, 133.68, 130.51, 127.30, 123.19, 120.14, 118.28, 118.04, 113.09, 110.74, 108.94, 107.48.

$^{19}$F NMR (376 MHz, CDCl$_3$) (δ): -61.78 (t, J = 26.4 Hz, 2F, O-C**F$_2$**-Ar), -110.39 (td, J = 26.4, 11.1 Hz, 2F, Ar-**F**), -113.87 (t, J = 9.7 Hz, 1F, Ar-**F**), -132.43 (dd, J = 20.9, 8.7 Hz, 2F, Ar-**F**), -163.10 (tt, J = 20.8, 4.8 Hz, 1F, Ar-**F**).

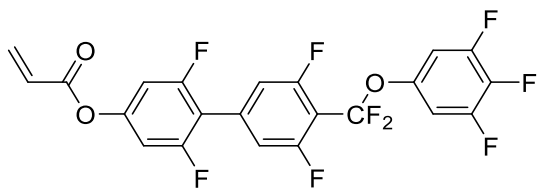

*4-acryloxy-2,6-difluoro-(3',5'-difluoro-4'-(difluoro(3,4,5-trifluorophenoxy)methyl)-1,1'-biphenyl)* **(3)**

Yield: (Colourless crystals) 365 mg, 74%

R$_F$ (50:50 Hexane:DCM): 0.55

$^1$H NMR (501 MHz, CDCl$_3$) (δ): 7.15 (d$_{apparent}$, J = 10.1 Hz, 2H, Ar-**F**), 7.00 (dd, J = 7.9, 5.8 Hz, 2H, Ar-**F**), 6.96 – 6.90 (m$_{apparent}$, 2H, Ar-**F**), 6.66 (dd, J = 17.3, 1.0 Hz, 1H, HC=C-**H**$_{trans}$H$_{Cis}$), 6.32 (dd, J = 17.3, 10.5 Hz, 1H, **H**C=CH$_{trans}$H$_{cis}$), 6.11 (dd, J = 10.5, 1.0 Hz, 1H, HC=C-H$_{trans}$**H**$_{Cis}$).

$^{13}$C{$^1$H} NMR (126 MHz, CDCl$_3$) (δ): 163.58, 161.04 – 158.65 (m)$^*$, 152.16 (ddd, J = 251.1, 10.7, 5.3 Hz), 151.87 (t, J = 14.4 Hz), 144.67 (td, J = 11.0, 1.7 Hz), 138.65 (dt, J = 250.4, 15.2 Hz), 134.51 (t, J = 11.5 Hz), 134.32, 127.11, 122.28, 120.16, 118.05, 115.25 – 114.59 (m$_{apparent}$), 113.05 (t, J = 17.9 Hz), 110.23 – 109.37 (m$_{apparent}$), 107.66 (dd, J = 18.7, 6.0 Hz), 106.72 (dd, J = 23.4, 6.9 Hz).

$^*$Overlapping signals

$^{13}$C{$^1$H}{$^{19}$F} NMR (126 MHz, CDCl$_3$) (δ): 163.44, 159.75, 159.74, 159.72, 159.67, 151.74, 144.53, 139.59, 134.37, 134.18, 126.98, 120.03, 114.74, 112.91, 109.67, 107.52, 106.59.

$^{19}$F NMR (400 MHz, CDCl$_3$) (δ): -61.95 (t, J = 26.5 Hz, 2F, O-C**F$_2$**-Ar), -110.59 (td, J = 26.6, 10.8 Hz, 2F, Ar-**F**), -112.06 (d, J = 9.3 Hz, 2F, Ar-**F**), -132.43 (dd, J = 20.7, 8.7 Hz, 2F, Ar-**F**), -163.06 (tt, J = 20.8, 6.0 Hz, 1F, Ar-**F**).

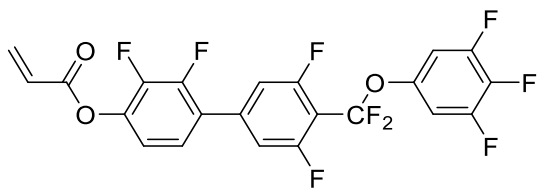

*4-acryloxy-2,3-difluoro-(3',5'-difluoro-4'-(difluoro(3,4,5-trifluorophenoxy)methyl)-1,1'-biphenyl)* **(4)**

Yield: (Colourless crystals) 278 mg, 56%

R$_F$ (50:50 Hexane:DCM): 0.56

$^1$H NMR (501 MHz, CDCl$_3$) (δ): 7.25 – 7.17 (m, 3H, Ar-**H**)*, 7.14 – 7.08 (m$_{apparent}$, 1H, Ar-**H**), 7.00 (dd, J = 7.9, 5.8 Hz, 2H, Ar-**H**), 6.70 (dd, J = 17.3, 1.0 Hz, 1H, HC=C-**H**$_{trans}$H$_{Cis}$), 6.38 (dd, J = 17.3, 10.5 Hz, 1H, **H**C=CH$_{trans}$H$_{cis}$), 6.14 (dd, J = 10.5, 1.0 Hz, 1H, HC=C-H$_{trans}$**H**$_{Cis}$).

$^{13}$C{$^1$H} NMR (126 MHz, CDCl$_3$) (δ): 163.01, 160.15 (dd, J = 258.6, 5.8 Hz), 151.17 (ddd, J = 251.1, 10.7, 5.2 Hz), 148.81 (dd, J = 253.8, 12.0 Hz), 144.98 (dd, J = 253.8, 14.8 Hz), 144.66 (td, J = 11.4, 4.5 Hz), 140.21 – 139.79 (m), 138.66 (dt, J = 250.6, 14.9 Hz), 134.60, 126.55, 125.22 (d, J = 9.9 Hz), 123.75 (t$_{apparent}$, J = 3.5 Hz), 122.29, 120.17, 119.29 (d, J = 4.0 Hz), 118.05, 113.27 (dd, J = 27.8, 3.3 Hz), 110.29 – 109.29 (m$_{apparent}$), 107.64 (dd, J = 18.5, 5.6 Hz).

$^{13}$C{$^1$H}{$^{19}$F} NMR (126 MHz, CDCl$_3$) (δ): 162.88, 160.02, 160.00, 144.53, 142.84, 139.94, 139.77, 139.55, 137.47, 134.46, 126.42, 125.08, 123.61, 120.03, 119.15, 113.13, 109.53, 107.50.

$^{19}$F NMR (376 MHz, CDCl$_3$) (δ): -61.86 (t, *J* = 26.4 Hz, 2F, O-C**F$_2$**-Ar ), -109.79 (td, *J* = 26.4, 10.8 Hz, 2F, Ar-**F**), -132.37 (dd, *J* = 20.7, 8.7 Hz, 2F, Ar-**F**), -138.86 (dd, *J* = 20.2, 7.4 Hz, 1F, Ar-**F**), -148.57 (dd, *J* = 20.3, 6.7 Hz, 1F, Ar-**F**), -162.99 (tt, *J* = 21.0, 6.0 Hz, 1F, Ar-**F**).

**NMR Spectra**

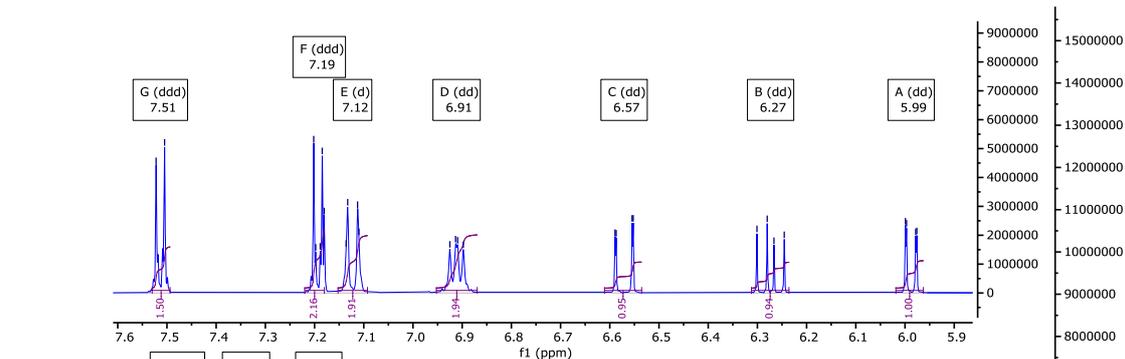
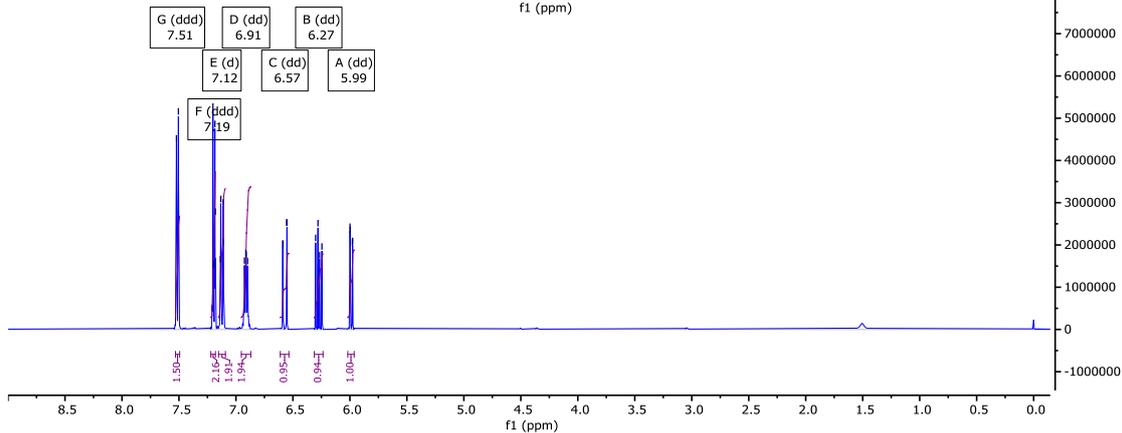
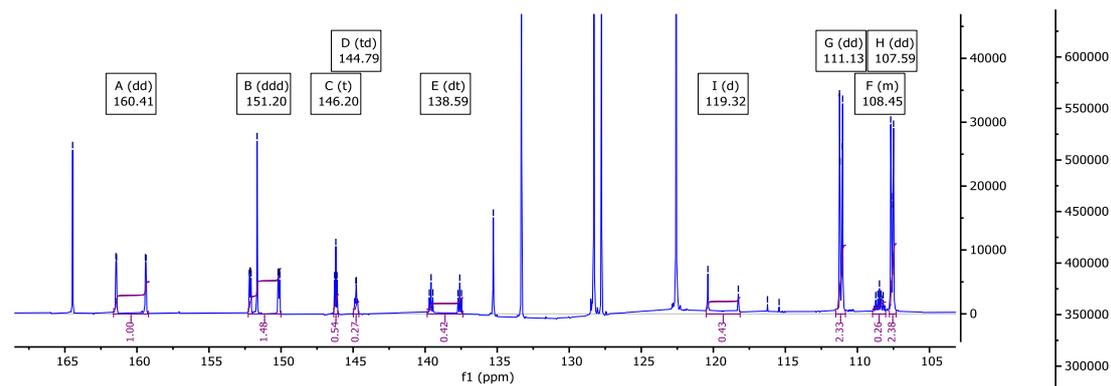
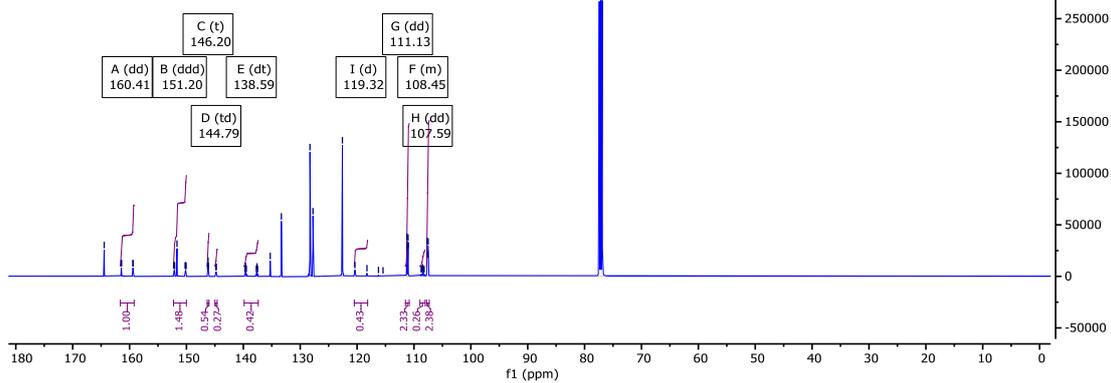

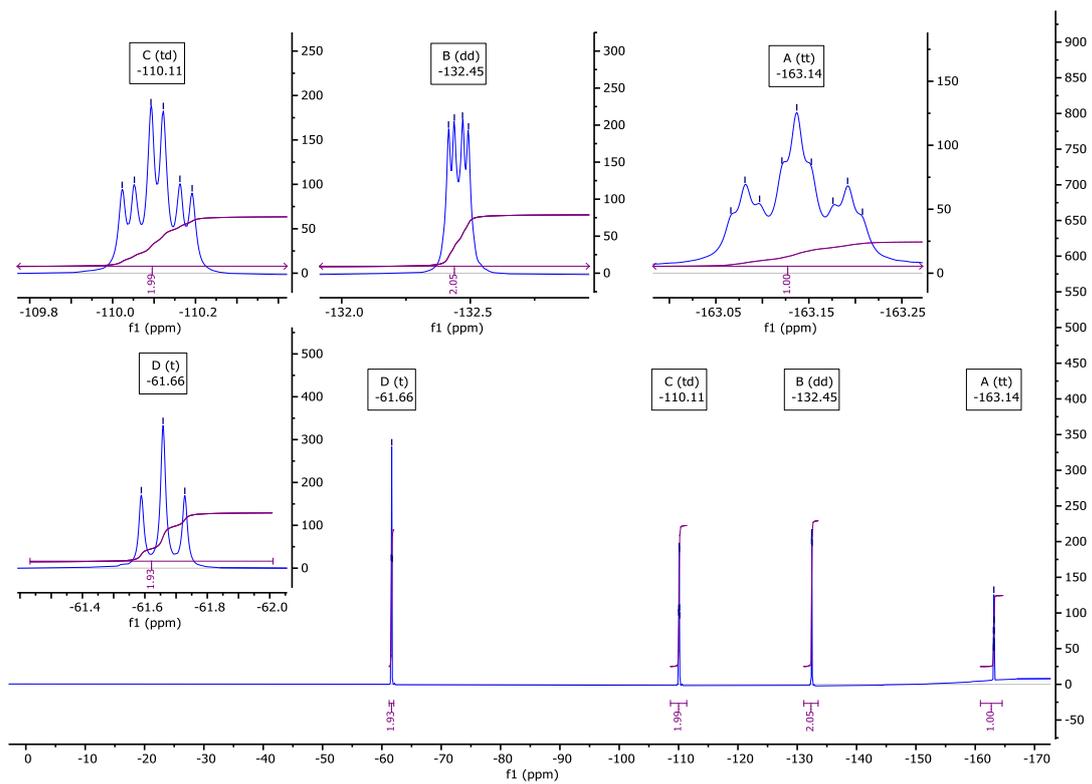

***Figure S1*** - *NMR spectra (In descending order: $^1$H, $^{13}$C{$^1$H}, and $^{19}$F, respectively), for* **(1)**.

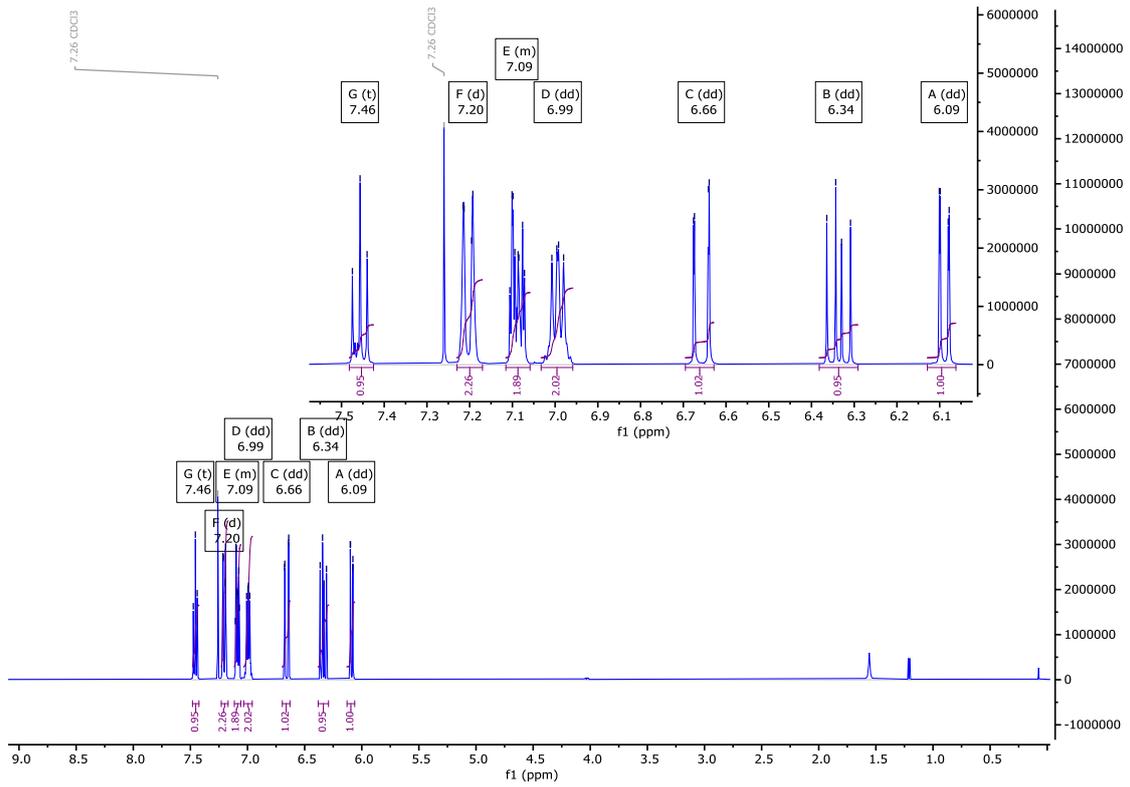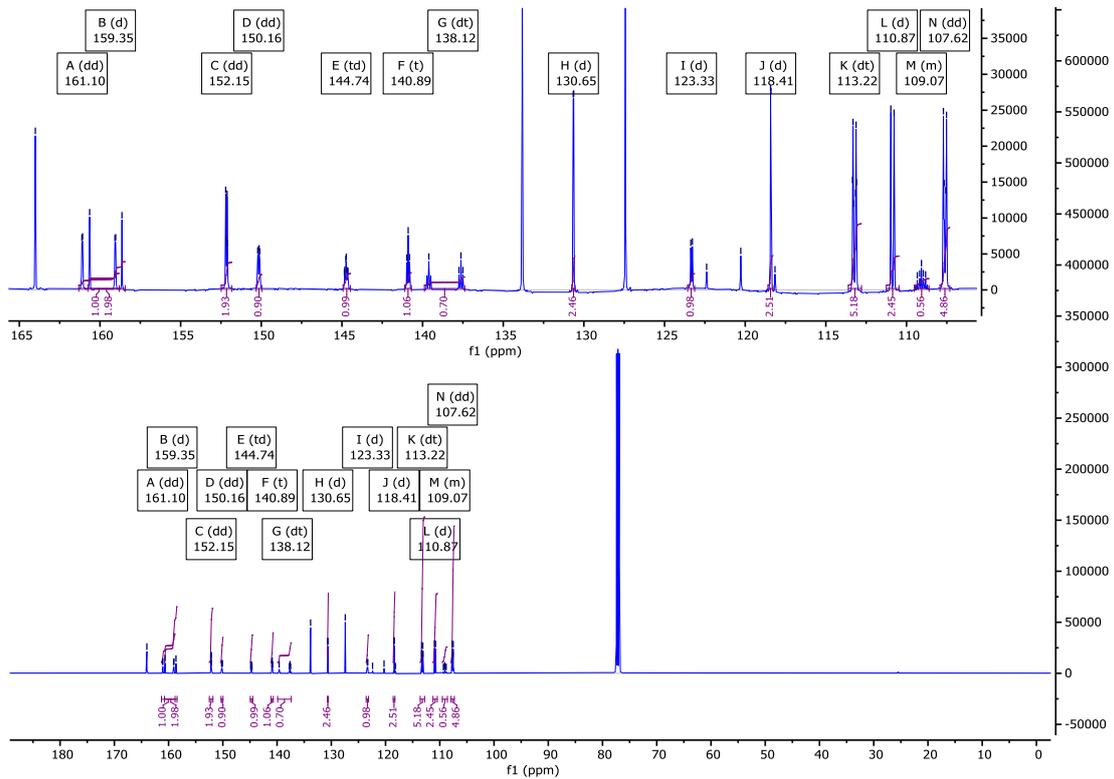

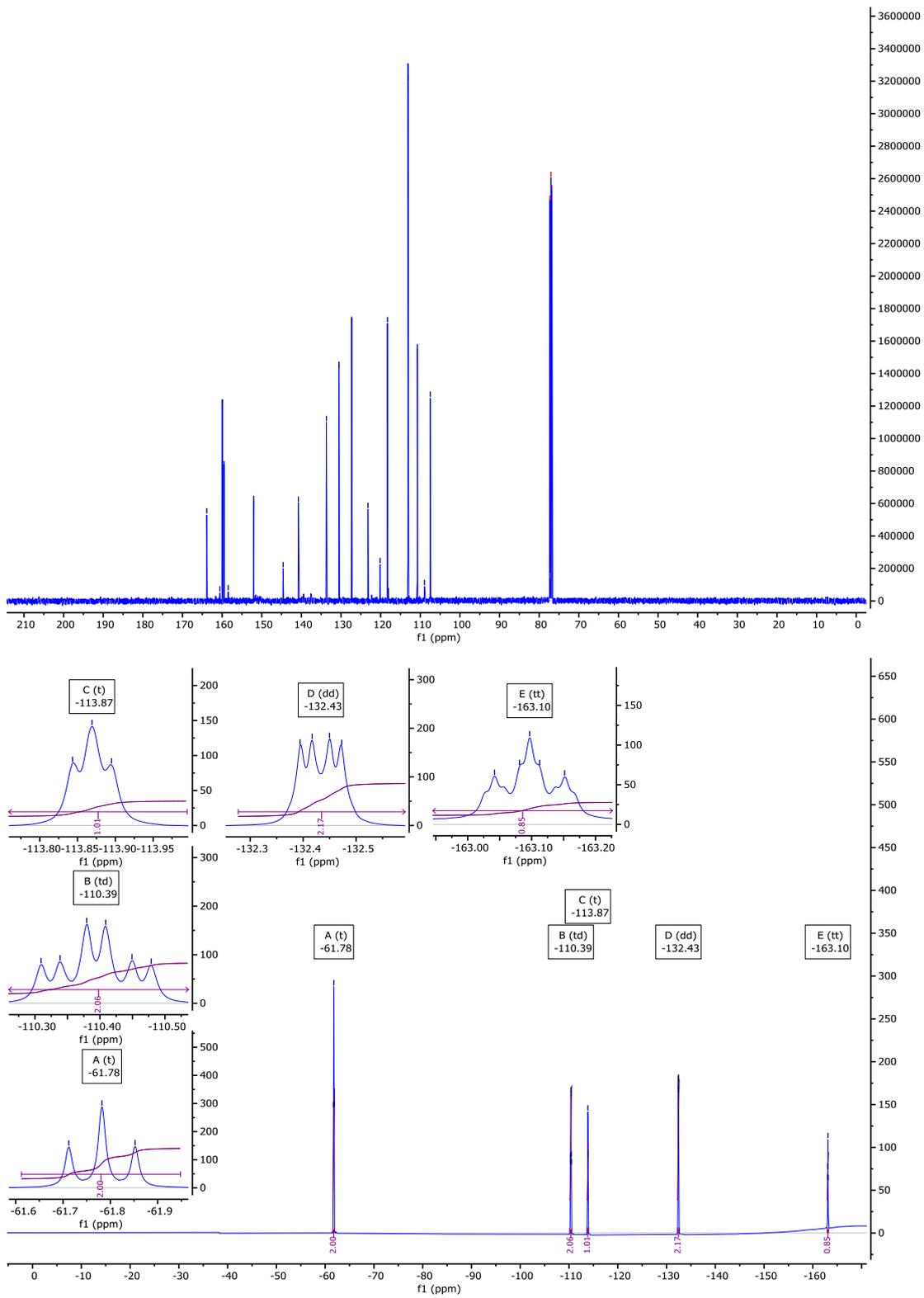

*Figure S2* - NMR spectra (In descending order: $^1H$, $^{13}C\{^1H\}$, $^{13}C\{^1H\}\{^{19}F\}$, and $^{19}F$, respectively) **(2)**.

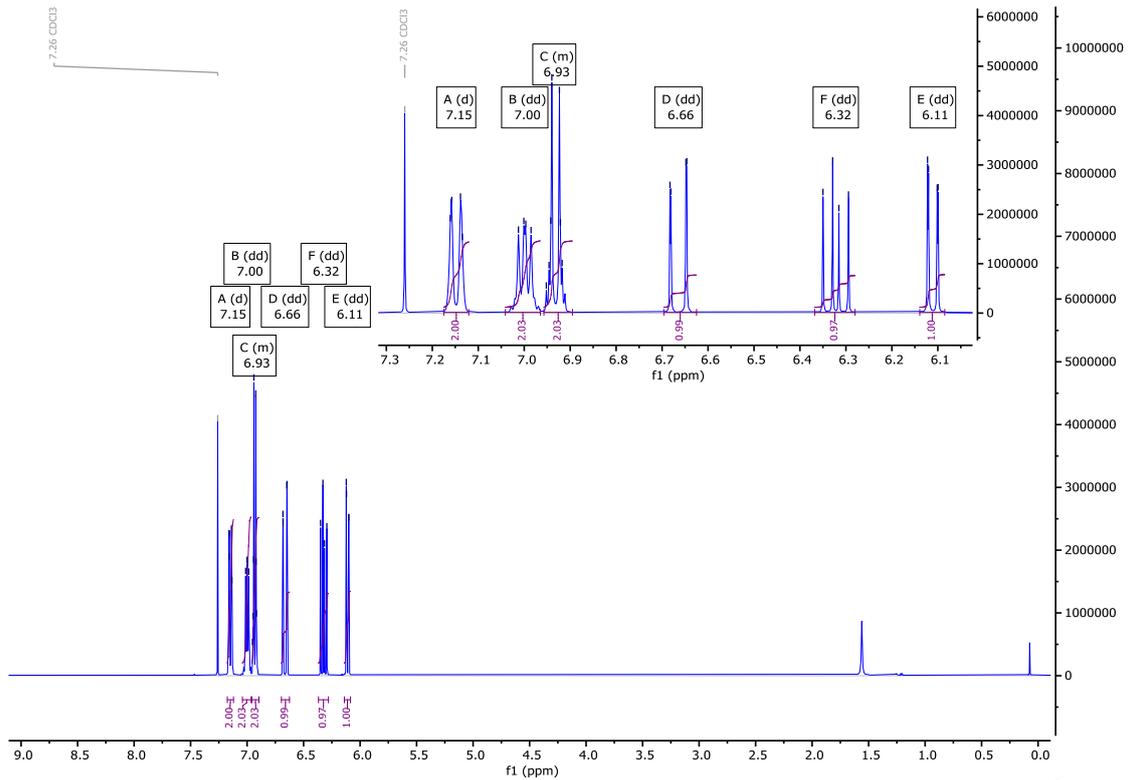
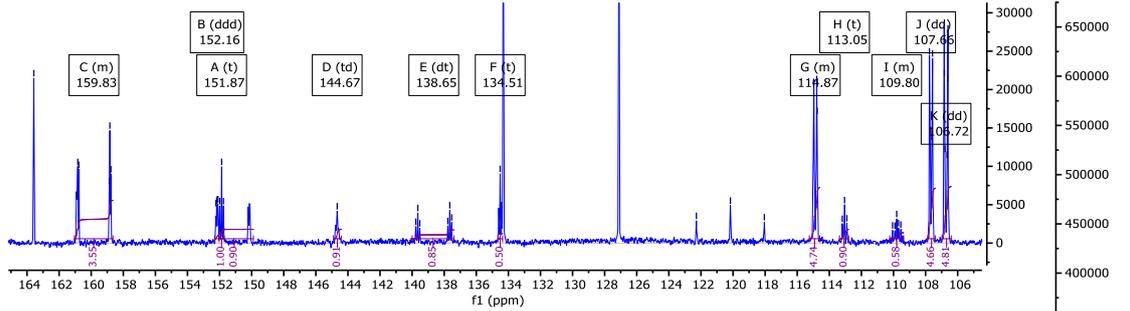
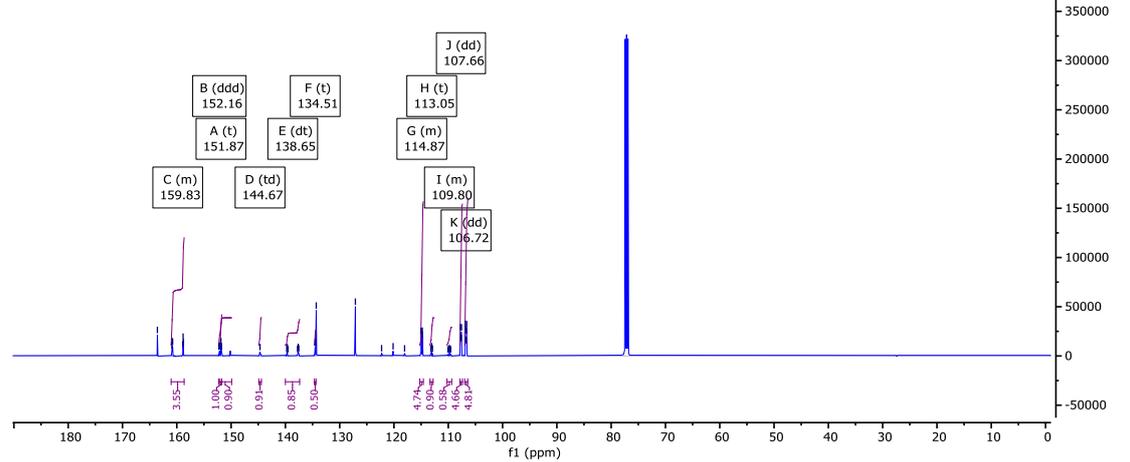

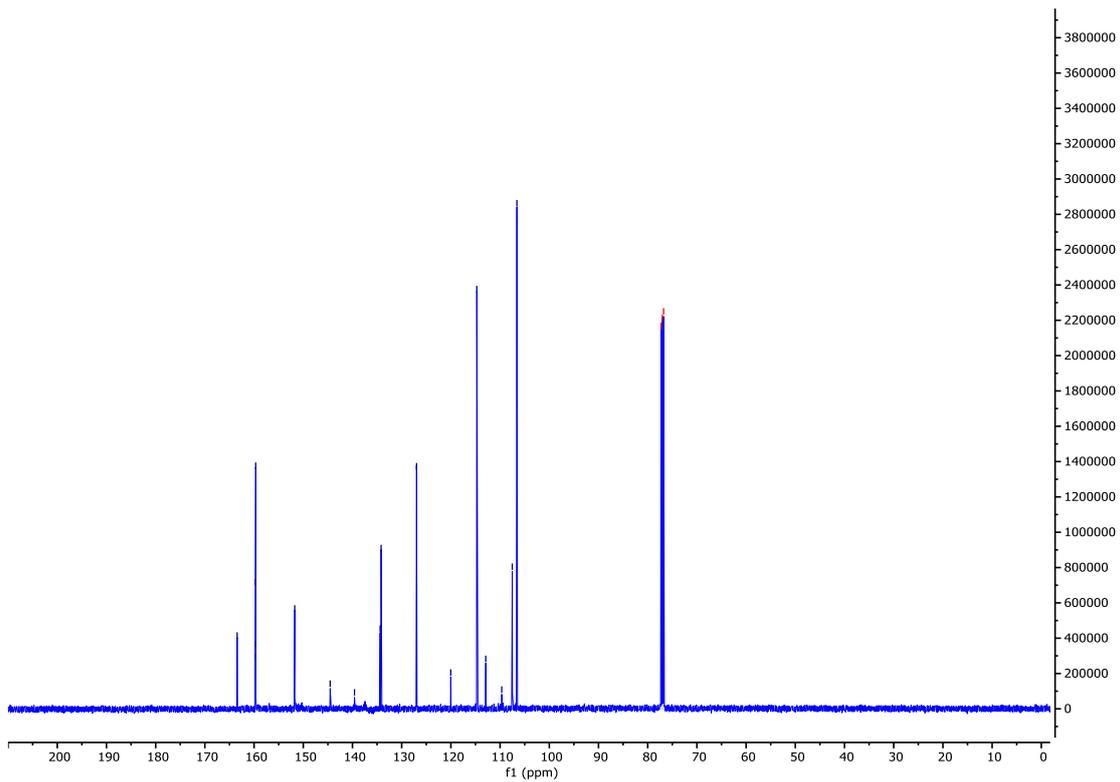
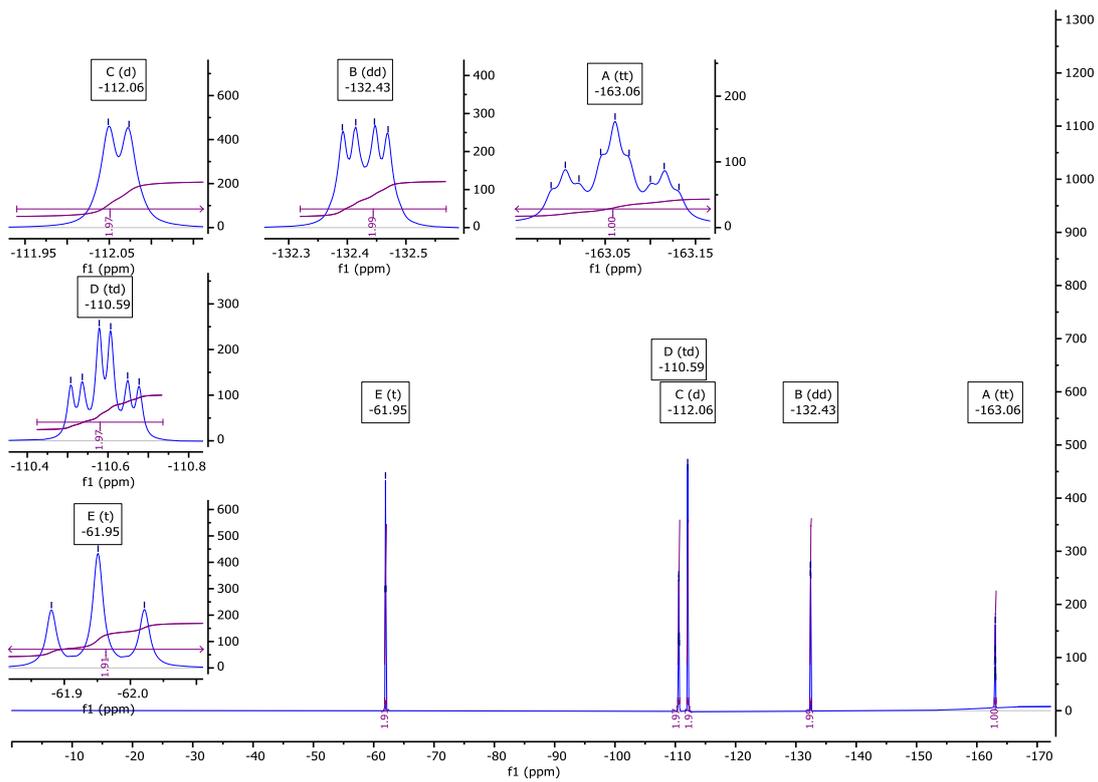

*Figure S3* - NMR spectra (In descending order: ¹H, ¹³C{¹H}, ¹³C{¹H}{¹⁹F}, and ¹⁹F, respectively) **(3)**.

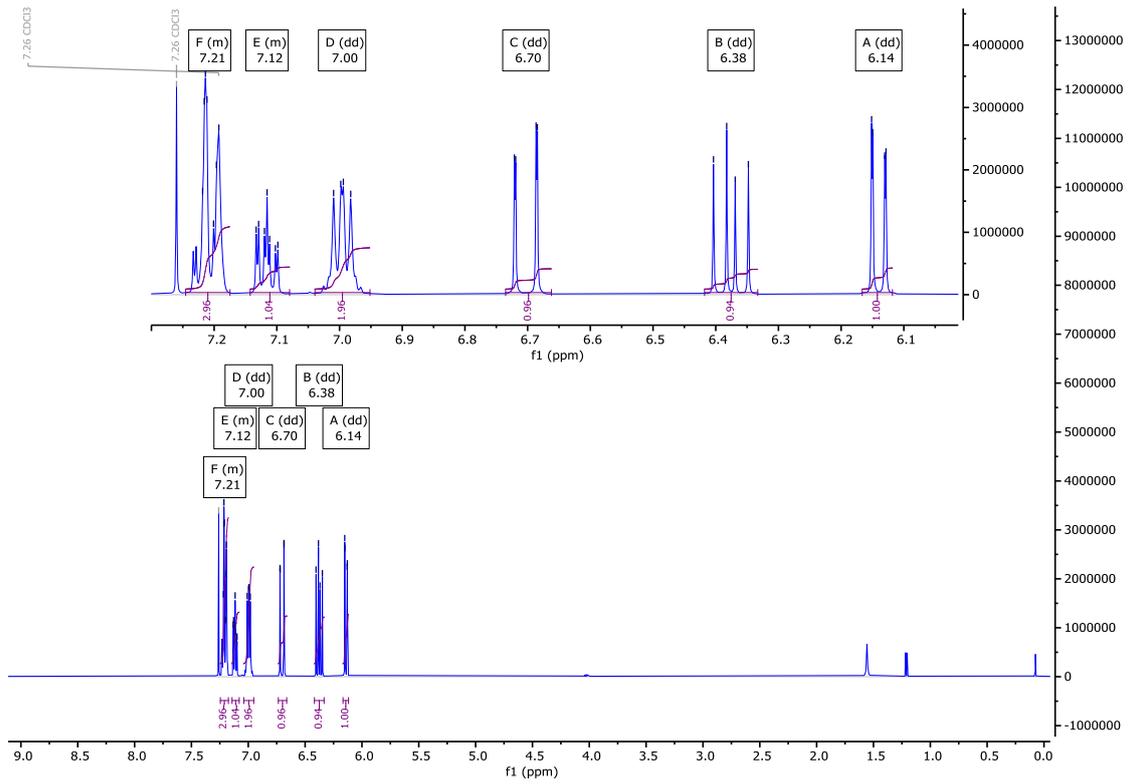
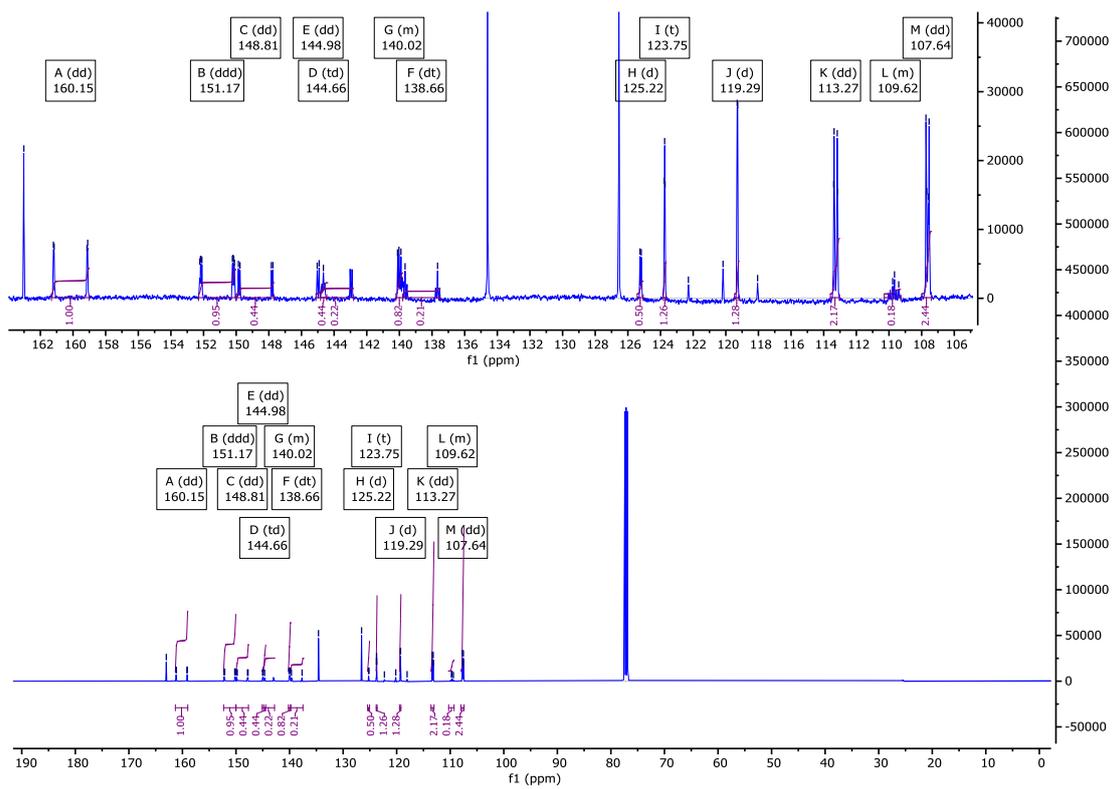

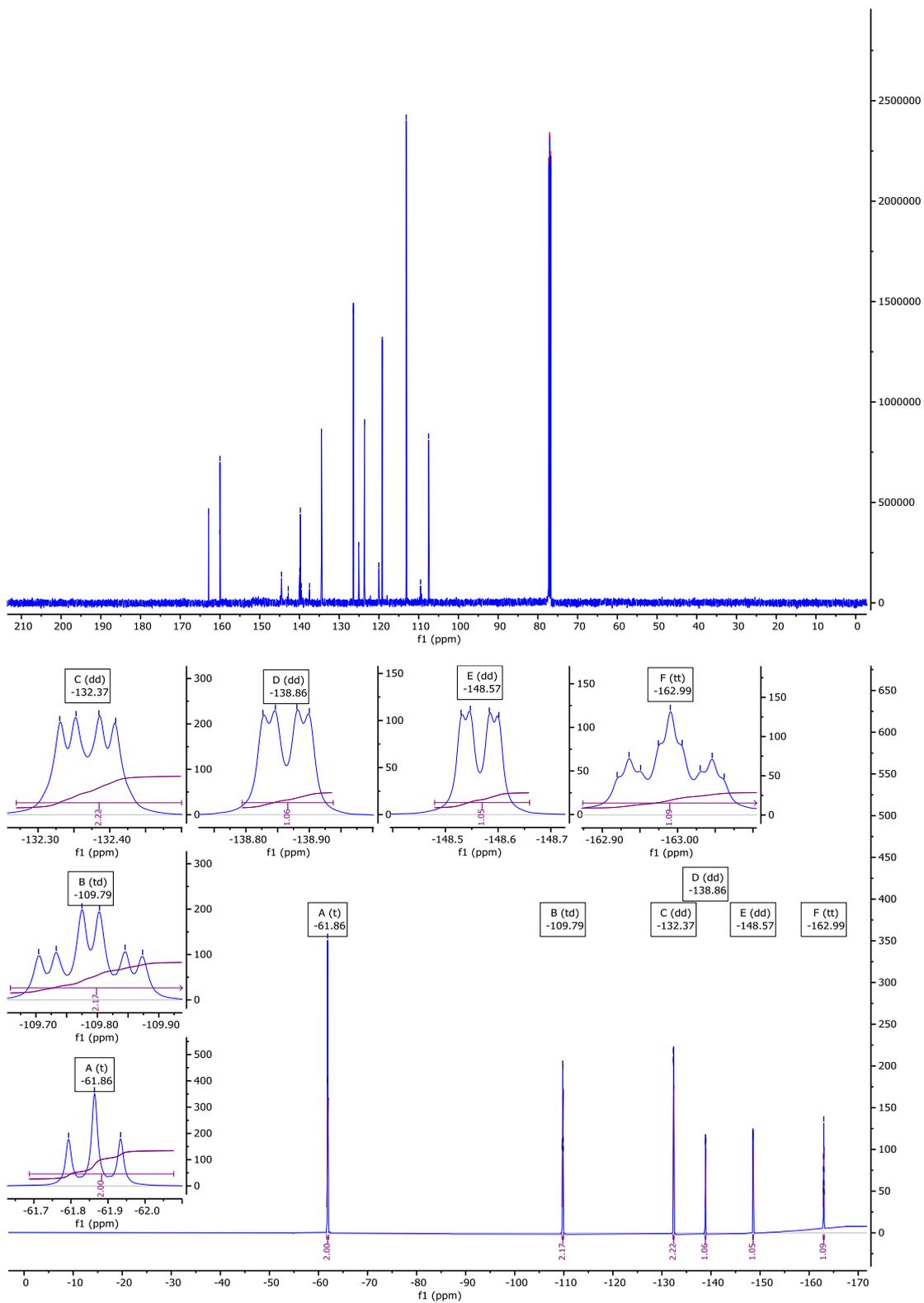

*Figure S4* - NMR spectra (In descending order: $^1H$, $^{13}C\{^1H\}$, $^{13}C\{^1H\}\{^{19}F\}$, and $^{19}F$, respectively) **(4)**.

## Characterisation of Monomers

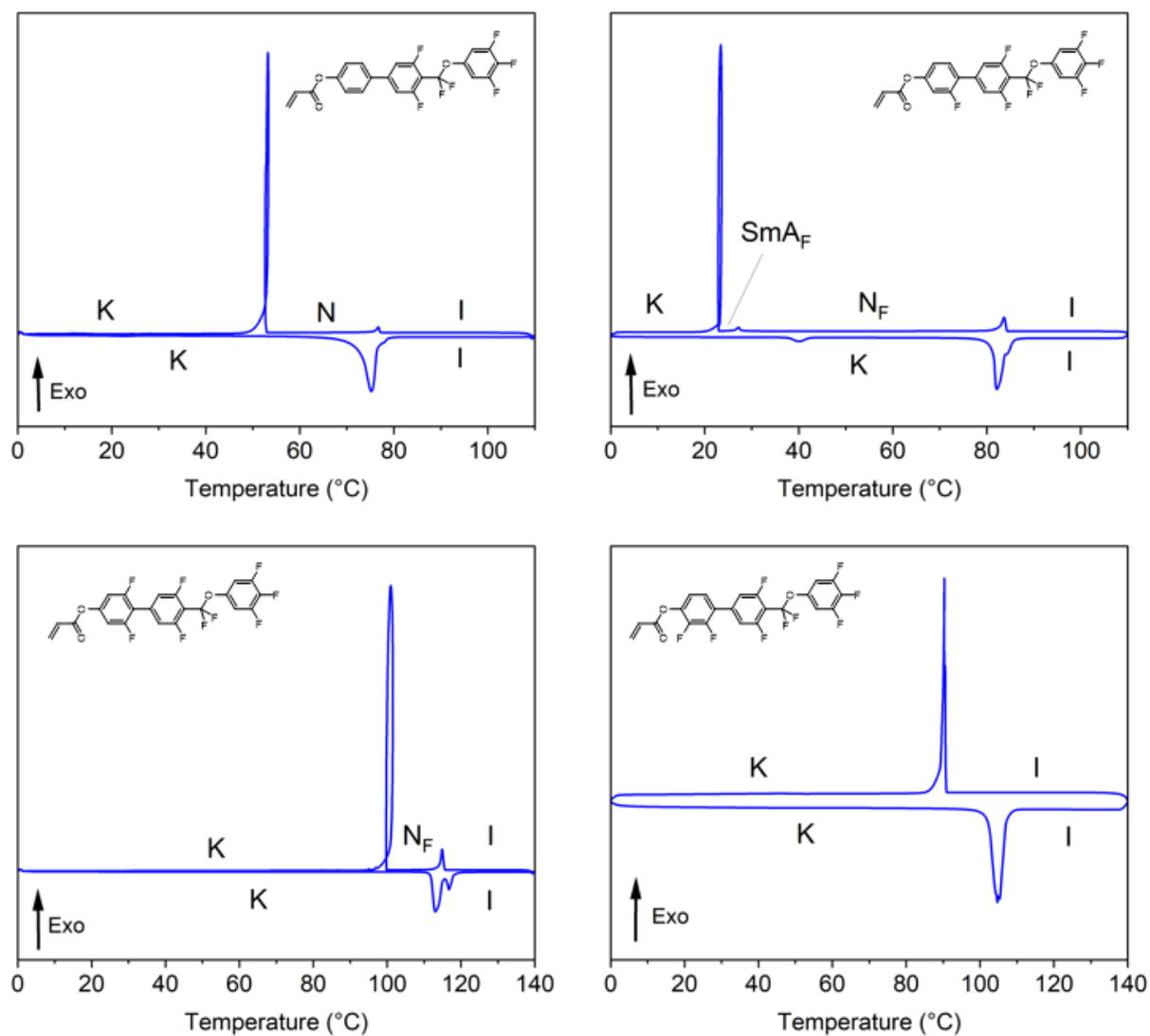

**Figure S5** – Example DSC thermograms for the synthesise monomers.

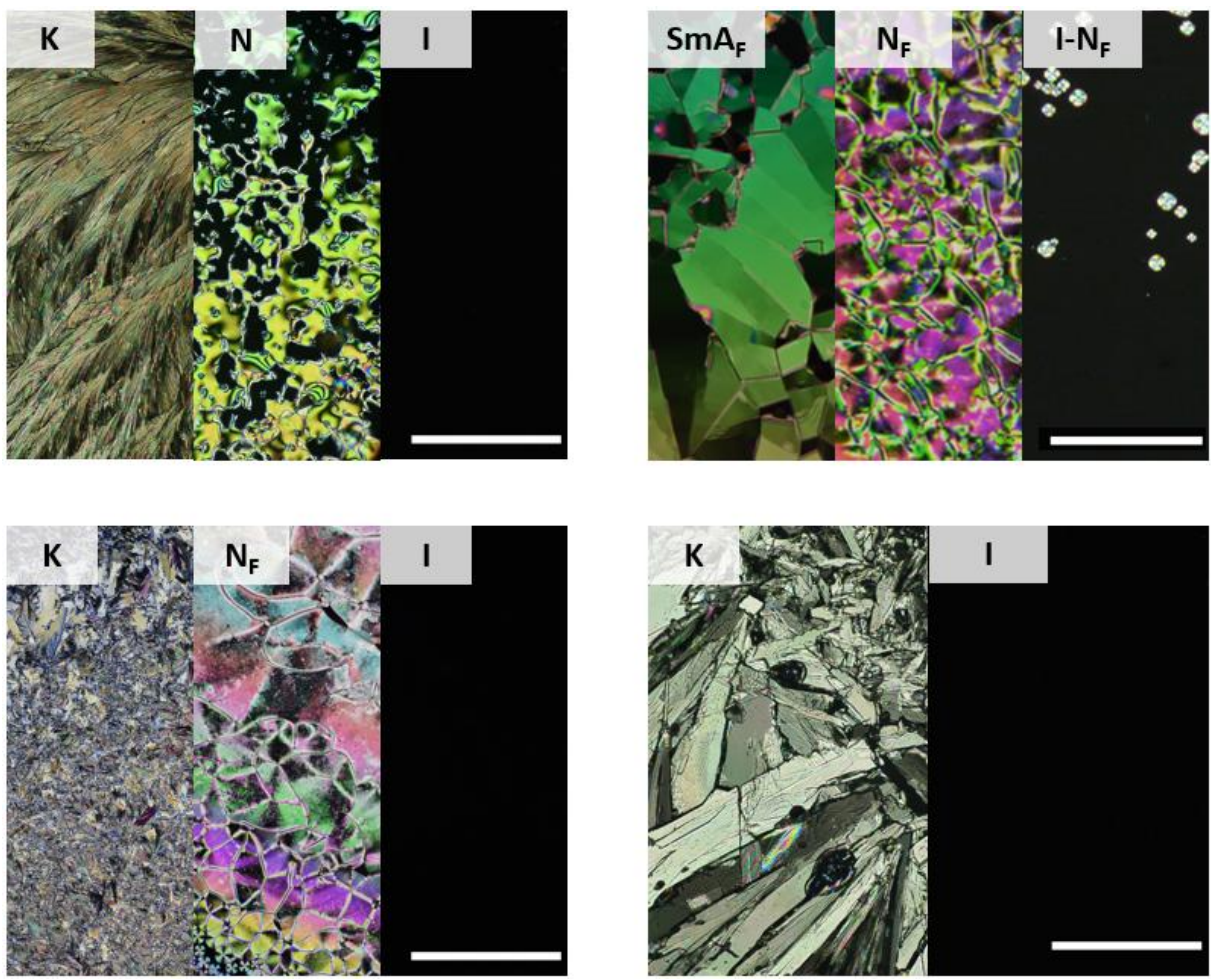

***Figure S6*** – *POM micrographs to show the phase behaviour for **1** (top left), **2** (top right), **3** (bottom left) and **4** (bottom right). In all cases the scale bar represents 500 µm.*

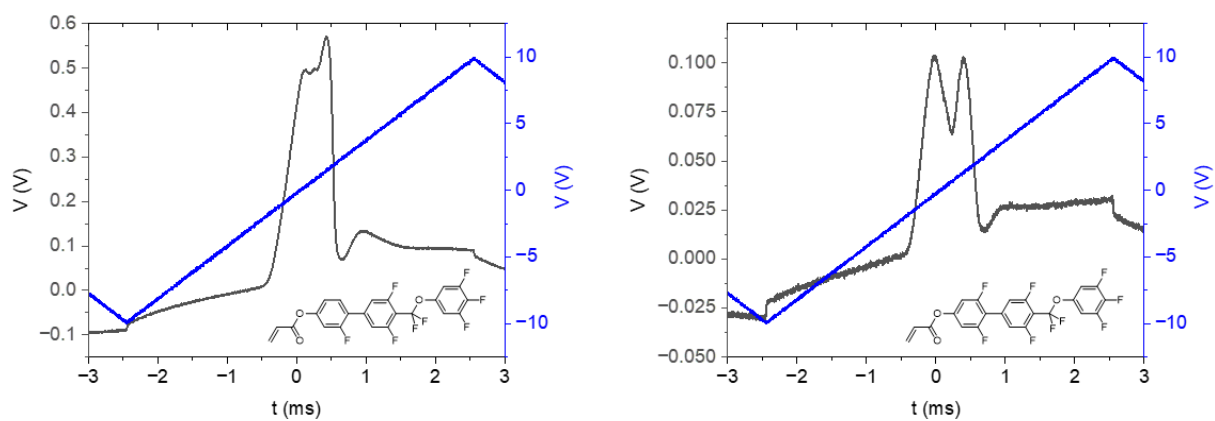

***Figure S7*** – *Current response traces measured for 2 and 3, confirming the assignment of the N$_F$ phase. In both cases measurements were taken at 100 Hz, at temperatures of 30 °C and 100 °C for monomers **2** and **3** respectively.*

## Polymer Stabilisation Studies

### Procedure

The appropriate quantities of **F7**, **RM82**, **2**, and **MBF** were added to DCM (2 mL), and the mixture stirred for 5 minutes. The solvent was then removed at elevated temperature to yield the desired mixture. The resulting mixture was sparged with nitrogen for 5 minutes before being filled into the appropriate sample container (e.g. liquid crystal cell). The mixtures were then subject to irradiation at 365 nm (2.5 W cm$^{-2}$) for 20 minutes whilst at room temperature, to yield the polymer stabilised mixture.

### Characterisation

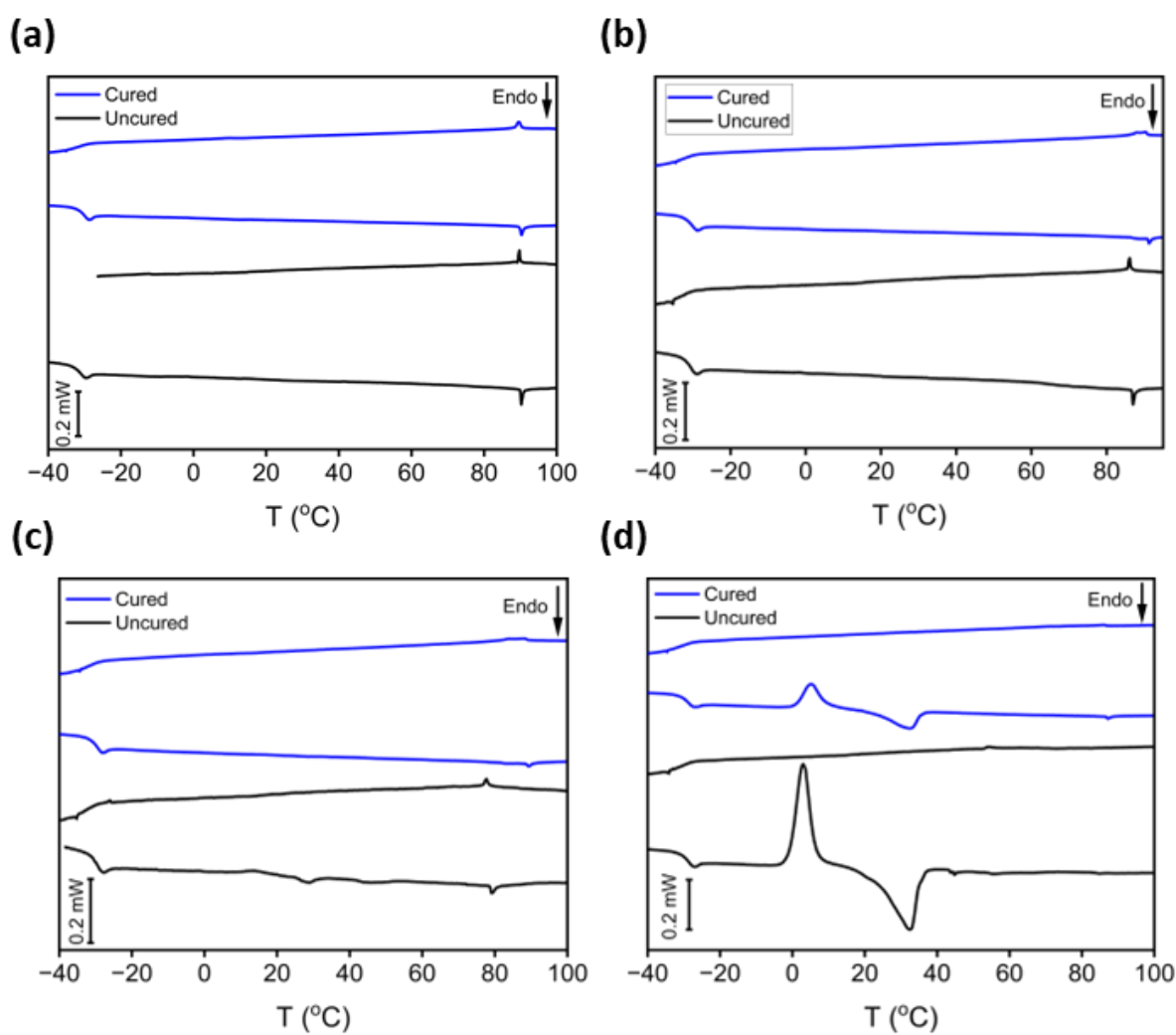

*Figure S8* – DSC thermograms for the polymer stabilisation experiments on mixtures PS1 (*(a)*), PS2 (*(b)*), PS3 (*(c)*) and PS4 (*(d)*). In all cases, the blue thermogram represents the cured mixture, and the black thermogram represents the uncured mixture.

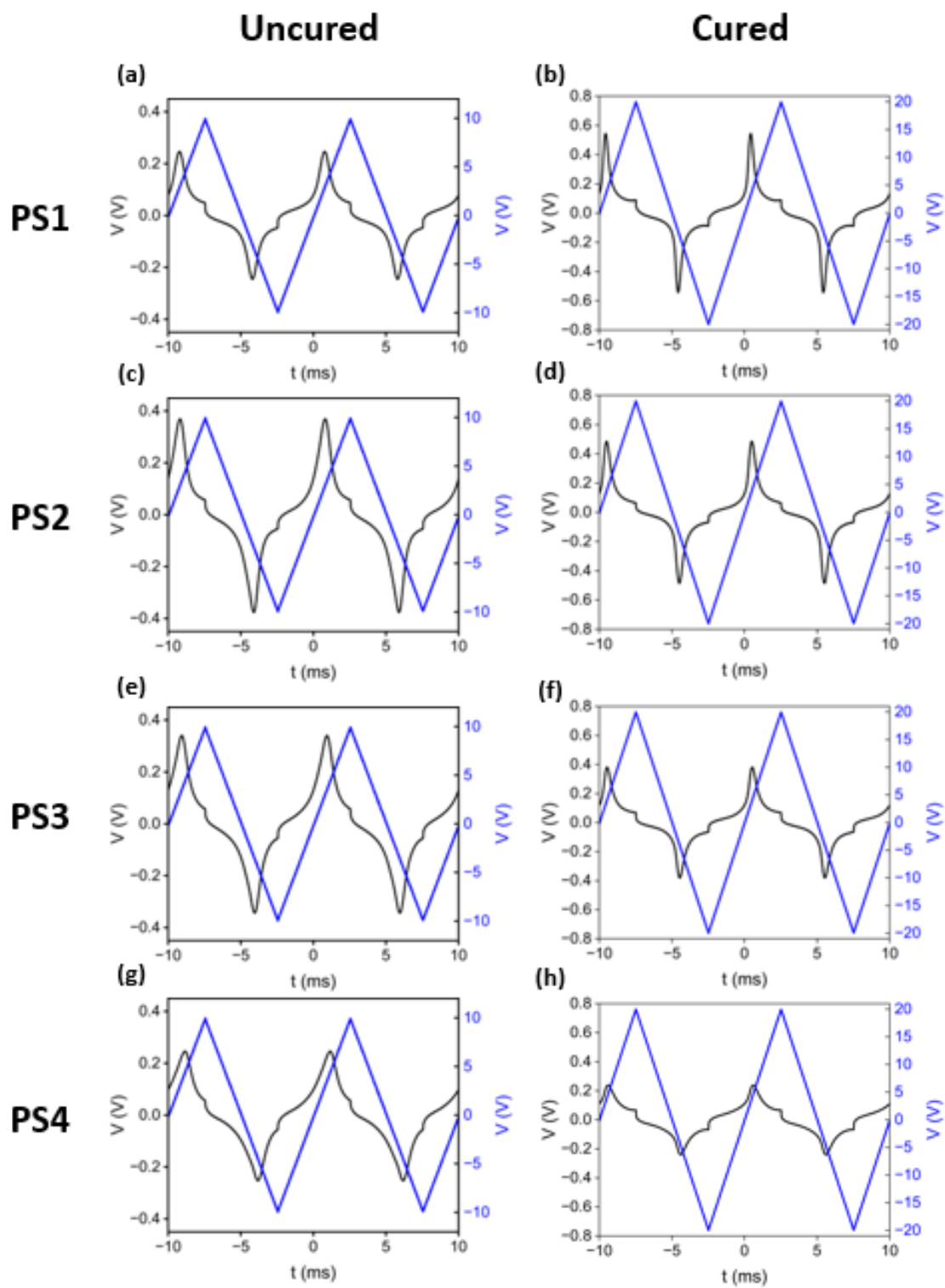

*Figure S9* – Current response traces for the polymer stabilisation mixtures. Panels *(a), (c), (e)* and *(g)* were acquired for uncured samples of **PS1, PS2, PS3** and **PS4** respectively. Panels *(b), (d), (f)* and *(h)* were obtained for cured samples of **PS1, PS2, PS3** and **PS4** respectively.